\documentclass{article}

\usepackage{arxiv}
\usepackage[utf8]{inputenc} 
\usepackage[T1]{fontenc}    
\usepackage{hyperref}       
\usepackage{url}            
\usepackage{booktabs}       
\usepackage{amsfonts}       
\usepackage{nicefrac}       
\usepackage{microtype}      
\usepackage{cleveref}       
\usepackage{lipsum}         
\usepackage{graphicx}
\usepackage[square,numbers]{natbib}
\usepackage{doi}
\usepackage{indentfirst}
\usepackage{xcolor}
\usepackage{caption}
\usepackage{subcaption}

\title{DeepTLS:  comprehensive and high-performance feature extraction for encrypted traffic}

\author{
	Zhi Liu \\
	Department of Computer Science\\
	Southwest Petroleum University\\
	Chengdu, China \\
	\texttt{zhi.liu@swpu.edu.cn} \\
}

\renewcommand{\headeright}{}
\renewcommand{\undertitle}{}

\hypersetup{
	pdftitle={deeptls},
	pdfauthor={Zhi Liu},
	pdfkeywords={Encrypted traffic, Feature extraction},
}

\begin{document}
\maketitle

\begin{abstract}

Feature extraction is critical for TLS traffic analysis using machine learning techniques, which it is also very difficult and time-consuming requiring huge engineering efforts. We designed and implemented DeepTLS,  a system which extracts full spectrum of features from pcaps across meta, statistical, SPLT, byte distribution, TLS header and certificates. The backend  is written in C++ to achieve high performance, which can analyze a GB-size pcap in a few minutes. DeepTLS was thoroughly evaluated against two state-of-the-art  tools Joy and Zeek with four well-known malicious traffic datasets consisted of 160 pcaps. Evaluation results show DeepTLS has advantage of analyzing large pcaps with half analysis time, and identified more certificates with acceptable performance loss compared with Joy. DeepTLS can significantly accelerate machine learning pipeline by reducing feature extraction time from hours even days to minutes. The system is online at \url{https://deeptls.com}, where test artifacts can be viewed and validated. In addition, two open source tools \href{https://github.com/zliucd/pysharkfeat}{Pysharkfeat} and \href{https://github.com/zliucd/tlsfeatmark}{Tlsfeatmark} are  also released.

\end{abstract}


\section{Introduction}

TLS is the mainstream encrypted protocol  widely used over Internet, accounting for around 90\% of visited websites\cite{enisareport}. In recent cyber attacks, nearly half of malware use TLS to conceal C\&C communication\cite{sophosreport} in order to bypass traditional defense systems such as firewall and IDS. 

Academic and industry have researched for  years to use machine learning  techniques, including deep learning, to combat malicious encrypted traffic. Feature extraction is critical to  build models, but it's also tremendously difficult and time-consuming for researchers and data scientists due to huge engineering efforts, which takes substantial cost and time in the pipeline.  

There are three essential capabilities of effective feature extraction: 1) support extensive features in all layers; 2) achieve high performance with instant analysis speed; 3) easy usability with minimum setup and use efforts. There have been several existing feature extraction tools, but they do not meet all these requirements.

Motivated to extract features efficiently and effectively with easier usability, we built the system DeepTLS. DeepTLS extracts all spectrum of features(70+) including meta, statistical, SPLT, byte distribution, TLS header and certificates.  The backend  is written in C++ to achieve high performance, which can analyze a GB-size pcap within a few minutes. In addition, DeepTLS is a web system with friendly interface, and as far as we know,  it is the first publicly available web system for dedicated feature extraction from TLS traffic.  

DeepTLS was thoroughly tested with two state-of-the-art tools Joy and Zeek using four well-known malicious traffic datasets consisted of 160 pcaps.  DeepTLS supports more features than Joy and Zeek, and test results show DeepTLS has advantage of analyzing large pcaps with half analysis time, and identified more certificates with acceptable performance loss than Joy.  Compared with Python implementation using Wireshark, DeepTLS is \textasciitilde 1000x faster. DeepTLS can be visited at \url{https://deeptls.com}, where test artifacts  can be viewed.  Meanwhile, to reproduce  testing of Joy, Zeek and Wireshark, two open source tools \href{https://github.com/zliucd/pysharkfeat}{Pysharkfeat} and \href{https://github.com/zliucd/tlsfeatmark}{Tlsfeatmark} are also released.

\section{Related work}

There are a number of research works that use machine learning to analyze TLS traffic for security analysis and traffic classification. \cite{rezaei2019deep} surveys encrypted traffic classification using deep learning. Cisco\cite{anderson2018deciphering} researchers proposed  to use SPLT, TLS header, certificate and other features to identify malicious TLS traffic, which achieves  high accuracy. \cite{diallo2021adaptive} uses header and statistical features to find malicious traffic. \cite{gomez2021unsupervised} combines TLS header, encrypted payload and certificate features to build unsupervised models. \cite{modi2019detecting} extracts several categories of features to find ransomware. \cite{barut2020machine} presents a detailed performance study on  encrypted traffic analysis. \cite{roques2019detecting, stvrasak2017detection} builds machine learning models to find anomalies in encrypted traffic. 

Feature extraction has been a crucial part to build models, and researches often use several tools such as Joy and Zeek to extract features from TLS traffic. Joy\cite{joy} is an open source feature extraction tool developed by Cisco researchers, which still requires certain post-processing of the output to build models. Zeek(formerly Bro\cite{paxson1999bro}) is a well-known network monitoring tool, which is widely used by research community to extract features especially TLS header and certificates, e.g., \cite{gomez2021unsupervised, roques2019detecting, stvrasak2017detection}. As a dedicated and state-of-the-art network protocol analyzer, Wireshark\cite{wireshark} can  also extract features. CICFlowmeter\cite{lashkari2017characterization} is desktop tool written in Java for feature extraction developed by University of New Brunswick.  Performance is important that Wireshark, Zeek and Joy are all written in low-level languages to achieve high performance. As Wireshark and Zeek do not provide native interface to extract features, researchers often write Python code to invoke their command lines and  parse the logs, which requires heavy post-processing and may incur high overhead(See Section 5.5).  Feature and capability comparison among different tools is described in more detail  in Section 3. 

In brief, feature extraction has been a vital but also time-consuming task. Our work is inspired  by current feature extraction tools described above to enable comprehensive and high performance feature extraction with easier usability.

\section{Feature and capability overview}

In this section, we give an overview of feature support and capability between DeepTLS and other major tools. As Joy\cite{anderson2018deciphering} proposed, features of encrypted TLS traffic can be divided into following categories. DeepTLS aims to extract all of them, which currently supports 70+ features  across these categories.

\begin{itemize}
	\item \textbf{Meta}: basic features of a TLS stream, e.g.,  source and dest IP addresses, source and dest ports, duration, stream index etc.
	\item \textbf{Statistical} uni-directional(inbound, outbound) and bi-directional(inbound and outbound) features of packet number, packet length, packet inter-arrival time and their sum/max/min/mean/std.
	\item \textbf{SPLT}: sequence of packet length and inter-arrival-time.
	\item \textbf{Byte distribution}: payload byte distribution, std and entropy of the distribution.
	\item  \textbf{TLS header}:  features derived from multiple TLS handshakes, e.g., TLS version, client cipher suites, server cipher suite, SNI.
	\item  \textbf{Certificate}: certificate version, serial,  valid time, issuer and subject information, public key type and length etc. 
\end{itemize}

\begin{table}[h]
\caption{Overview of feature and capability comparison}
\centering
	\begin{tabular}	{p{2.5cm} p{2.5cm} p{2.5cm}  p{2.5cm} p{2.5cm}}
		\toprule
	&  \textbf{Wireshark}   &  \textbf{Zeek} &  \textbf{Joy} & \textbf{DeepTLS}  \\
		\midrule
		Meta feats& yes  & yes & yes & \textbf{yes}     \\
		Statistical  feats  & partial & partial & partial & \textbf{yes(30+)}      \\
		SPLT  feats   & no & no & yes & \textbf{yes}     \\
		Byte dist. feats& no & no  & yes & \textbf{yes} \\
		TLS feats& full & yes & yes & \textbf{yes} \\
		X509 feats & full & yes & yes & \textbf{yes} \\
		JSON output & yes & yes & yes & \textbf{yes} \\
		Post processing & heavy & heavy & some &  \textbf{little} \\
		Usage & parse logs &  parse logs  & source code & \textbf{web} \\
		Mode & offline & offline & realtime, offline & \textbf{offline} \\
		Protocol & many & many & a few & \textbf{TLS only} \\
		\bottomrule
	\end{tabular}
\end{table}

Table 1 shows a brief comparison of feature support and other capabilities between DeepTLS and other major tools; a more detailed comparison can be found in {\url{https://deeptls.com/features}. In meta category, Zeek and Joy do not extract stream index \footnote{Stream index is useful in Wireshark to locate TCP stream, and Zeek has a similar field \texttt{uid} in conn.log to identify unique connections. DeepTLS adds stream index for each TCP stream using same mechanism as Wireshark.}, which is important to cross-check examine features and manual analysis with  DPI tools(e.g., Wireshark). In statistical category, DeepTLS supports full statistical features(30+) and other tools support a small subset of them.  In SPLT and byte distribution category, only Joy\footnote{\texttt{SRLT}(Sequence of Record Lengths and Times) stands for SPLT in JSON output of Joy.}   and DeepTLS have support.  In TLS header and certificate category, all tools can extract  multiple features from TLS handshakes and certificates. As the state-of-the-art protocol decoder, Wireshark has premium TLS decoding capability, and DeepTLS extracts some important features from several TLS handshakes, e.g.,  \texttt{Client Hello, Server Hello, Certificates}.  All tools support JSON output, but the outputs of Wireshark and Zeek require heavy post-processing to generate features. Though the JSON output of Joy is human-readable, it still requires some efforts to process.  DeepTLS produces intuitive JSON output, and samples of two TLS streams can be viewed at {\url{https://deeptls.com/samples}}. 

Two major differences between DeepTLS and other tools are usage and mode. In usage,  DeepTLS is web-based, and Joy provides source code for local compilation and command line use. DeepTLS works offline which analyzes pcaps and does not support realtime analysis currently; Joy support both realtime and offline analysis.   Though Wireshark and Zeek can work in realtime and offline, but for feature extraction they work only in offline by parsing  logs to generate features. Also, DeepTLS only analyzes TLS traffic, and Joy can analyze a few other protocols(e.g., DNS and HTTP) in addition to TLS \footnote{https://github.com/cisco/joy, see Overview section.};  Wireshark and Zeek, as state-of-the-art DPI tools, have rich protocol decoding capability supporting a variety of protocols in addition to TLS.

\section{Design and implementation}
\subsection{Architecture overview}

DeepTLS is composed of backend and frontend. The backend is written in C++ to achieve high performance, which uses open source library for layer3 and layer4 packet parsing. We have developed a dedicated TLS decoder \textbf{Dissector} to accurately identify TLS handshakes and analyze the semantics. The frontend is a web server that takes features generated by backend and exposes web access.

\begin{figure}[h]
\centering
\includegraphics[scale=0.42]{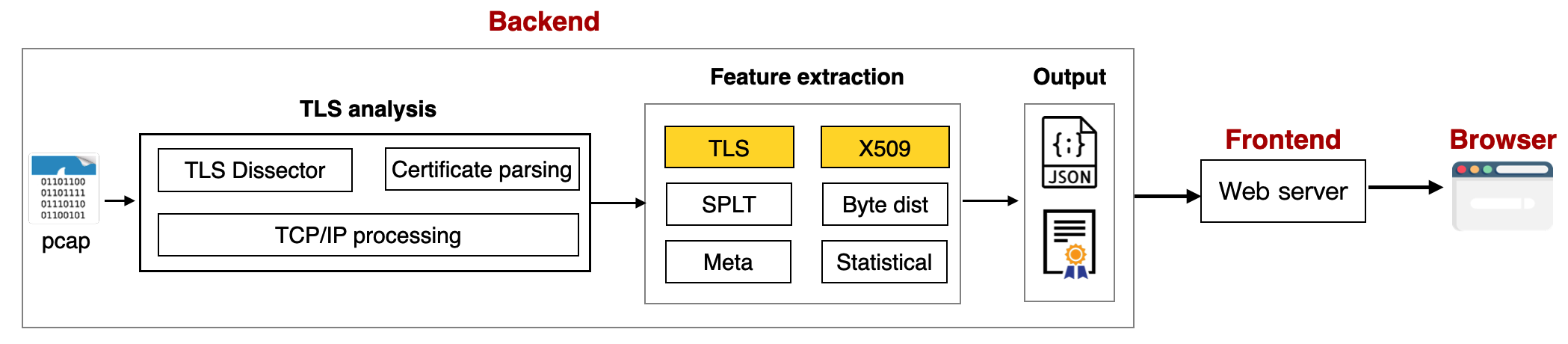}
\caption{DeepTLS architecture}
\end{figure}

\subsection{TLS Dissector}

Dissector  identifies TLS handshakes and analyze their protocol semantics, including \texttt{Client Hello, Server Hello, Certificates} and other messages. Besides, DeepTLS is able to handle several complex scenarios that are commonly used in TLS transmission especially malicious TLS traffic.

\begin{itemize}
\item \textbf{Reassembly}: certificates are sent in multiple consecutive packets following \texttt{Client Hello}, and receiver reassemble these packets to reconstruct full data.
\item \textbf{Multi-handshake messages}: several handshake messages share one record header.
\end{itemize}

\subsection{Certificate parsing}

X509 certificates are difficult to parse due to complex ANS.1 encoding.  DeepTLS integrates an industry-level crypto library mbedtls\cite{mbedtls} to parse certificates that ensures accuracy and robustness.

\section{Evaluation}

\subsection{Setup}

DeepTLS was evaluated in three dimensions against two state-of-the-art tools Joy and Zeek: performance, the number of TLS streams identified and certificates parsed(including certificate duplicates, see footnote 7). Each test was run for five times, and analysis time in tables is the average time of each analysis task. In the following tables, in addition to  the whole dataset, several pcaps which spend longer analysis time or produce more TLS streams are also listed, and the suffix ".pcap" of the file name except ISCX2014 is ignored for brevity\footnote{To view pcap name of tested pcaps, visit  \url{https://deeptls.com/datasets}, and click specific dataset.}.

To make the testing solid and transparent, artifacts analyzed by DeepTLS can be viewed and validated in \url{https://deeptls.com/artifacts}. We developed an open source tool \href{https://github.com/zliucd/tlsfeatmark}{Tlsfeatmark}  to reproduce  testing of Joy and Zeek, and test results can be found in the \href{https://github.com/zliucd/tlsfeatmark/tree/master/output/testing_four_datasets}{repository} of Tlsfeatmark. 

The testing environment is Centos 8.5 64-bit running on a physical machine with CPU i7-11700K(8 cores) 3.60GHz, 32GB RAM and built-in gcc/g++ 8.5.

\subsection{Datasets}

Four well-known and widely-used datasets by research community are chosen as benchmarks having 160 pcaps, which contain real normal and malicious TLS traffic from multiple malware families.

\begin{itemize}
\item  \href{https://www.unb.ca/cic/datasets/botnet.html}{ISCX2014}:  2 large pcaps(5GB and 2.1GB) with botnet TLS traffic  \footnote{The third pcap is derived from \texttt{testing.pcap} and have duplicate TLS streams,  which is ignored to analyze.}
\item \href{https://github.com/ojroques/tls-malware-detection}{Malware-stratosphere} (or Stratosphere in this paper): 30 pcaps with malicious traffic generated by multiple malware families
\item \href{https://www.malware-traffic-analysis.net/2021/index.html}{MTA2021}: 115 pcaps from \href{https//www.malware-traffic-analysis.net}{Malware Traffic Analysis} site collected in 2021
\item \href{https://www.stratosphereips.org/datasets-ctu13}{CTU13}: 13 pcaps with normal and botnet traffic
\end{itemize}

\subsection{Testing tools}

\begin{itemize}
	\item DeepTLS
	\item Tlsfeatmark(commit \texttt{  \href{https://github.com/zliucd/tlsfeatmark/tree/2535f62122e13aae6f2cda629a711e16d5d112d4}{2535f62}}): the open source tool written in Python to reproduce testing of Joy and Zeek.
	\begin{itemize}
		\item   Joy(version 4.5.0, major release version) \\
		 \texttt{joy tls=1 dist=1 bidir=1 fpx=1 entropy=1 model=F1:F2  xx.pcap | gunzip}
	\end{itemize}
	\begin{itemize}
		\item   Zeek(version 4.2.0, major release version) \\
		 \texttt{zeek -r xx.pcap}
	\end{itemize}
	\item Pysharkfeat(commit \texttt{\href{https://github.com/zliucd/pysharkfeat/tree/c0c3e77c43878517286bafc7cc570b3cd7048b6d}{c0c3e77}}): Python implementation of feature extraction using Wireshark(version 3.0.0 released in 2019\footnote{In newer versions of Wireshark such as 3.4.x and upward, \texttt{tshark} is much slower than older versions. }), currently supporting meta, statistical, SPLT and byte distribution.

\end{itemize}

\subsection{Results}

Table 2 shows test results in four datasets on three dimensions: performance, the number of TLS streams identified and certificated parsed \footnote{Duplicate certifcates are taken into account, and the unique certifcate number for each tool is much smaller(see Section 5.4.3).} In all datasets, Zeek has shortest analysis time and extracted the largest number of certificates. DeepTLS is 1.5x slower than Zeek and Joy but identified more certificates than Joy. Joy has marginal performance loss compared with  Zeek, and identified close number of TLS streams with Zeek and certificates with DeepTLS. 

\begin{table}[h!]
\caption{Evaluation results of four datasets} 
\centering
\begin{tabular}{p{2cm} p{2.2cm} p{2.2cm} p{2.2cm} p{1.5cm} p{2.2cm}}
	\toprule
	& Time(seconds)  & TLS streams   & Certificates      \\ 
	\midrule
	Joy   & 746.2          &  724645         & 947783          \\
	Zeek             & 711.7        & 711780        &  1169839          \\
	\textbf{DeepTLS} & \textbf{1069.3} & \textbf{533243} & \textbf{966346}  \\
	\bottomrule
\end{tabular}
\end{table}

\subsubsection{Performance analysis}

In  four datasets, Joy has overall better performance in Stratosphere and MTA2021, so Joy is the baseline. In ISCX2014 with two large GB-size pcaps, DeepTLS outperformed Joy  with over half analysis time.  In other two datasets Stratosphere and MTA2021 containing large number of TLS and certificates, DeepTLS is up to \textasciitilde2.6x slower than the baseline. 

\begin{table}[h!]
\caption{Analysis time  of ISCX2014} 
\centering
\begin{tabular}{p{3cm} p{2.5cm} p{2.5cm}  p{2.2cm} p{2.4cm}}
	\toprule
	& training.pcap      & testing.pcap       & Total         & Overhead      \\ 
	\midrule
	Joy & 124.5s         & 80.3s          & 204.8s         & 1.00x(baseline)           \\
	Zeek          & 118.5s         & 65.7s          & 184.2s         & 0.9x           \\
	\textbf{DeepTLS}       & \textbf{64.0s} & \textbf{30.4s} & \textbf{97.3s} & \textbf{0.48x}  \\
	\bottomrule
\end{tabular}
\end{table}

\begin{table}[h!]
\caption{Analysis time  of Malware-stratosphere} 
\centering
\begin{tabular}{p{2cm} p{2.2cm} p{2.2cm} p{2.2cm} p{1.5cm} p{2.2cm}}
	\toprule
	& vawtrak\_0001 & trickbot\_0014 & trickbot\_0021 & Total          & Overhead      \\ 
	\midrule
	Joy & 35.9s          & 16.9s           & 13.1s           & 143.3 s         & 1.00x(baseline)          \\
	Zeek          & 51.1s          & 27.4s           & 20.0 s          & 218.2s          & 1.52x           \\
	\textbf{DeepTLS}       & \textbf{45.4s} & \textbf{30.2s}  & \textbf{23.7s}  & \textbf{376.4s} & \textbf{2.63x}  \\
	\bottomrule
\end{tabular}
\end{table}

\begin{table}[h!]
\caption{Analysis time  of MTA2021} 
\centering
\begin{tabular}{p{2cm} p{2.2cm} p{2.2cm} p{2.2cm} p{1.5cm} p{2.2cm}}
	\toprule
	&  12-13-Contact &  09-23-Gozi-2 &  10-01-Qakbot &  Total &  Overhead \\
	\midrule
	Joy &            1.6s &           1.3s &           0.7s &   29.7s &       1.00x(baseline) \\
	Zeek &            2.1s &           2.0s &           0.7s &   42.2s &       1.42x \\
	\textbf{DeepTLS}       & \textbf{2.7s} & \textbf{2.1s}  & \textbf{1.1s}  & \textbf{41.1s} & \textbf{1.38x}  \\    	
	\bottomrule
\end{tabular}
\end{table}

\begin{table}[h!]
\caption{Analysis time  of CTU13} 
\centering
\begin{tabular}{p{2cm} p{2.2cm} p{2.2cm} p{2.2cm} p{1.5cm} p{2.2cm}}
	\toprule
	& fast-flux-2  & 0810-neris   & 0817-bot      & Total          & Overhead      \\ 
	\midrule
	Joy   & 2.0s          & 1.3s          & 10.5s          & 368.4 s         & 1.00x(baseline)           \\
	Zeek             & 4.3s          & 2.8s          & 16.4s          & 267.1s          & 0.73 x          \\
	\textbf{DeepTLS} & \textbf{4.6s} & \textbf{2.0s} & \textbf{20.4s} & \textbf{554.5s} & \textbf{1.51x}  \\
	\bottomrule
\end{tabular}
\end{table}

\subsubsection{TLS streams analysis}

As Zeek is widely used to extract certificate and TLS  header features in research community, Zeek is the baseline in TLS stream and certificate evaluation. All three tools identified a number of TLS streams, but Zeek surprisingly missed a considerable number of TLS streams in ISCX2014(only 267, see Table 7).  In Stratosphere,  DeepTLS identified fewer TLS streams(0.73x than Joy and Zeek), but it extracted more certificates than Joy(929368 and 919732 respectively, see Table 12).  As certificates are derived from TLS handshakes,  it indicates DeepTLS is able to find more valid TLS streams.  In MTA2021, three tools found almost same number of TLS streams(see Table9).  


\begin{table}[h!]
\caption{TLS streams identified in ISCX2014}
\centering
\begin{tabular}{p{3cm} p{2.5cm} p{2.5cm}  p{2.2cm} p{2.4cm}}
	\toprule
	& training.pcap      & testing.pcap       & Total         & Benchmark        \\ 
	\midrule
	Joy              & 5199          & 2644          & 7843          & 29.37x           \\
	Zeek             & 135           & 132           & 267           & 1.00x(baseline)  \\
	\textbf{DeepTLS} & \textbf{2902} & \textbf{1152} & \textbf{4054} & \textbf{15.18x}  \\
	\bottomrule
\end{tabular}
\end{table}

\begin{table}[h!]
\caption{TLS streams identified in Malware-stratosphere}
\centering
\begin{tabular}{p{2cm} p{2.2cm} p{2.2cm} p{2.2cm} p{1.5cm} p{2.2cm}}
	\toprule
	& vawtrak\_0001  & trickbot\_0014 & trickbot\_0021 & Total           & Benchmark       \\ 
	\midrule
	Joy              & 144411         & 94507          & 103823         & 682965          & 1.01x           \\
	Zeek   & 142901         & 94507          & 103822         & 675689          & 1.00x(baseline)           \\
	\textbf{DeepTLS} & \textbf{49223} & \textbf{68451} & \textbf{80133} & \textbf{493712} & \textbf{0.73x}  \\
	\bottomrule
\end{tabular}
\end{table}

\begin{table}[h!]
\caption{TLS streams identified in MTA2021}
\centering
\begin{tabular}{p{2cm} p{2.2cm} p{2.2cm} p{2.2cm} p{1.5cm} p{2.2cm}}
	\toprule
	& 12-13-Contact & 09-23-Gozi-2  & 10-01-Qakbot & Total          & Benchmark        \\ 
	\midrule
	Joy              & 7539          & 3060          & 500          & 32333          & 1.00x            \\
	Zeek             & 7539          & 3054          & 715          & 32190          & 1.00x(baseline)  \\
	\textbf{DeepTLS} & \textbf{7539} & \textbf{3076} & \textbf{714} & \textbf{32995} & \textbf{1.03x}   \\
	\bottomrule
\end{tabular}
\end{table}

\begin{table}[h!]
\caption{TLS streams identified in CTU13}
\centering
\begin{tabular}{p{2cm} p{2.2cm} p{2.2cm} p{2.2cm} p{1.5cm} p{2.2cm}}
	\toprule
	& fast-flux-2  & 0810-neris  & 0817-bot      & Total         & Benchmark         \\ 
	\midrule
	Joy              & 94           & 33          & 1347          & 1504          & 0.41x             \\
	Zeek             & 370          & 63          & 3086          & 3634          & 1.00x(baseline)  \\
	\textbf{DeepTLS} & \textbf{171} & \textbf{60} & \textbf{2199} & \textbf{2482} & \textbf{0.68x}    \\
	\bottomrule
\end{tabular}
\end{table}

\subsubsection{Certificates analysis}

As certificates are derived from TLS handshakes, the number of certificate is a reasonable indicator of TLS decoding capability.  In ISCX2014, Zeek only parsed 510 certificates, and the other two tools parsed a lot more certificates, and the number of certificates parsed by DeepTLS(4922) is  as twice as Joy(2307), and over 9x than Zeek(510). In three other datasets, Zeek extracted the largest number of certificates, which shows Zeek has excellent certificate extraction capability.

It's interesting to see there are many duplicates of  certificates parsed that among one million certificates found and parsed by DeepTLS(as well other tools), there are only 2305 unique certificates based on  binary hash. One possible reason is there many identical domain or IP visits via TLS for both benign and malicious traffic across these pcaps.


\begin{table}[h!]
\caption{Certificates parsed in ISCX2014}
\centering
\begin{tabular}{p{3cm} p{2.5cm} p{2.5cm}  p{2.2cm} p{2.4cm}}
	\toprule
	& training.pcap      & testing.pcap       & Total         & Benchmark        \\ 
	\midrule
	Joy              & 1661          & 646           & 2307          & 4.52x            \\
	Zeek             & 330           & 180           & 510           & 1.00x(baseline)  \\
	\textbf{DeepTLS} & \textbf{3522} & \textbf{1400} & \textbf{4922} & \textbf{9.65x}   \\
	\bottomrule
\end{tabular}
\end{table}

\begin{table}[h!]
\caption{Certificates parsed in Malware-stratosphere}
\centering
\begin{tabular}{p{2cm} p{2.2cm} p{2.2cm} p{2.2cm} p{1.5cm} p{2.2cm}}
	\toprule
	& vawtrak\_0001   & trickbot\_0014  & trickbot\_0021 & Total           & Benchmark        \\ 
	\midrule
	Joy              & 284236          & 118673          & 79556          & 919732          & 0.83x            \\
	Zeek(baseline)   & 285800          & 153165          & 143599         & 1113524         & 1.00x(baseline)  \\
	\textbf{DeepTLS} & \textbf{285804} & \textbf{118674} & \textbf{79573} & \textbf{929368} & \textbf{0.83x}   \\
	\bottomrule
\end{tabular}
\end{table}

\begin{table}[h!]
\caption{Certificates parsed in MTA2021}
\centering
\begin{tabular}{p{2cm} p{2.2cm} p{2.2cm} p{2.2cm} p{1.5cm} p{2.2cm}}
	\toprule
	& 12-13-Contact & 09-23-Gozi-2  & 10-01-Qakbot  & Total          & Benchmark        \\ 
	\midrule
	Joy              & 581           & 9080          & 1018          & 23813          & 0.51x            \\
	Zeek   & 7723          & 9083          & 1593          & 46401          & 1.00x(baseline)  \\
	\textbf{DeepTLS} & \textbf{586}  & \textbf{9111} & \textbf{1439} & \textbf{25407} & \textbf{0.55x}   \\
	\bottomrule
\end{tabular}
\end{table}

\begin{table}[h!]
\caption{Certificates parsed in CTU13}
\centering
\begin{tabular}{p{2cm} p{2.2cm} p{2.2cm} p{2.2cm} p{1.5cm} p{2.2cm}}
	\toprule
	& fast-flux-2  & 0810-neris   & 0817-bot      & Total         & Benchmark        \\ 
	\midrule
	Joy              & 138          & 61           & 1669          & 1931          & 0.21x            \\
	Zeek   & 966          & 134          & 8017          & 9404          & 1.00x(baseline)  \\
	\textbf{DeepTLS} & \textbf{468} & \textbf{121} & \textbf{5928} & \textbf{6649} & \textbf{0.71x}   \\
	\bottomrule
\end{tabular}
\end{table}

\subsection{Wireshark-based implementation test}

Analyzing logs of Wireshark is another approach to extract features, which may incur very high overhead. To test the performance of DeepTLS against this approach, we developed a tool Pysharkfeat which extracts features by invoking Wireshark command line \texttt{tshark} and parsing the logs, and Table 15 shows results on two benchmarks MTA2021-01-02 and \texttt{training.pcap} from ISCX2014.  MTA2021-01-02 has 20 pcaps from MTA2021 collected from January to February in 2021.   For \texttt{training.pcap}, Pysharkfeat was running over 24 hours(actually the analysis did not stop over 30 hours in our test), and we stopped Pysharkfeat after that period of time, and use 24 hours as the analysis time. Test results show Pysharkfeat incurs exceedingly high overhead, and DeepTLS is \textasciitilde 1000x faster.

Wireshark is fast, but invoking \texttt{tshark} to parse each TCP stream is orders of magnitude slower than native Wireshark.  This result indicates it's not practical in engineering perspective to use this approach to extract features in time-critical tasks. The major difference of post-processing between Wireshark and Zeek is that Zeek generates a bunch of logs for a pcap at one time, and Wireshark-based feature extraction finds TCP streams indexes first and then analyze each stream one by one \footnote{More details can be found in the source code of Pysharkfeat and Tlsfeatmark.}.

\begin{table}[h!]
\caption{Performance evaluation of Pysharkfeat}
	\begin{subtable}[h]{0.45\textwidth}
		\centering
		\caption{Analysis time of MTA2021-01-02}
		\begin{tabular}{p{2cm} p{2cm} p{2cm}}
			\toprule
			Tool   &  Time   & Overhead  \\
			\midrule
			Pysharkfeat & 3329.6s & 832x \\
			\textbf{DeepTLS} & \textbf{4.0s} & \textbf{1x(baseline)} \\
			\bottomrule
		\end{tabular}
	\end{subtable}
	\hfill
	\begin{subtable}[h]{0.45\textwidth}
		\caption{Analysis time of training.pcap from ISCX2014}
		\begin{tabular}{p{2cm} p{2cm} p{2cm}}
			\toprule
			Tool   &  Time& Overhead  \\
			\midrule
			Pysharkfeat & >24 hours & 1319x \\
			\textbf{DeepTLS} &\textbf{ 65.5s} & \textbf{1x(baseline)} \\
			\bottomrule
		\end{tabular}
	\end{subtable}
\end{table}

\section{Conclusion}

Feature extraction has been a major obstacle to analyze malicious TLS traffic using machine learning. 
We built the system DeepTLS to extract comprehensive features from TLS traffic fast with friendly interface.  DeepTLS was tested with two major feature extraction tools Joy and Zeek using several well-known datasets. Test artifacts of DeepTLS are public and can be browsed in \url{https://deeptls.com/artifacts}.  With our released tools Pysharkfeat and Tlsfeatmark, it is convenient to reproduce testing of Joy, Zeek and Wireshark.   

We are going to release labeled malicious TLS datasets, which could help researchers and data scientists build and evaluate machine learning models more effectively.

\bibliographystyle{unsrt}
\bibliography{deeptls}

\end{document}